\documentclass[conference,10pt,letterpaper]{IEEEtran}

\usepackage{amsmath,amssymb,amsfonts}
\usepackage{algorithmic}
\usepackage{graphicx}
\usepackage{textcomp}
\usepackage{xcolor}
\usepackage{float}
\usepackage{caption}
\usepackage{subcaption}
\usepackage[utf8]{inputenc}
\usepackage{mathtools, nccmath}
\usepackage[LGRgreek]{mathastext}
\usepackage[percent]{overpic}
\usepackage{upgreek}
\usepackage{lipsum,capt-of,graphicx}

\usepackage[font={footnotesize}]{caption}

\usepackage[margin=0.63in]{geometry}

\begin{document}

\title{\fontsize{22}{22}\selectfont{\textit{Ab initio} Modeling of MoS$_2$/Oxide Device Interfaces with Machine Learned Electronic Structures}}

\author{\fontsize{11}{11}\selectfont M. Kaniselvan, M. Dossena, D. Lu, A. Maeder, N. Vetsch, A. N. Ziogas, and M. Luisier\\
\fontsize{10}{12}\selectfont Integrated Systems Laboratory, ETH Zurich, Zurich, Switzerland, email: mkaniselvan@iis.ee.ethz.ch}

\maketitle


\noindent \textbf{\textit{Abstract}}--- We introduce a new \textit{ab initio} approach to simulate semiconductor devices that integrates scalable machine-learned (ML) electronic structure models with an advanced quantum transport (QT) solver. The developed framework enables 10,000$\times$ speedups over density functional theory to produce the Hamiltonian matrix of devices made of $>$20,000 atoms, while offering high prediction accuracy. We use its unique features to investigate MoS$_2$/oxide samples and single-layer MoS$_2$ field-effect transistors, where the surrounding oxide layers, here, HfO$_2$ or Al$_2$O$_3$, are explicitly included into the QT domain. In particular, we reveal that the presence of undercoordinated metal atoms (Hf or Al) close to the semiconductor-oxide interface significantly affects the magnitude of the electronic current and its propagation through MoS$_2$. 

\section{\textbf{Introduction}}

Significant progress has been made towards the miniaturization of field-effect transistors (FETs) based on mono- and few-layer transition metal dichalcogenide (TMDC). Several research groups have demonstrated, for example, MoS$_2$ devices with an equivalent oxide thickness EOT=0.8~nm~[1], a channel length $L_g$=30~nm~[2], or a width $W$=25~nm~[3]. Moreover, a record ON-state current $I_{ON}$=1,135~$\mu$A/$\mu$m at $V_{ds}$=1.5 V was reported for monolayer MoS$_2$~[4]. While impressive, this number falls short of theoretical predictions: \textit{ab initio} quantum transport (QT) simulations suggest that MoS$_2$ could deliver $I_{ON}$ exceeding 2,000~$\mu$A/$\mu$m. Ref.~[5] even anticipates $I_{ON}>$3,000~$\mu$A/$\mu$m at $V_{ds}$=0.64~V, accounting for contact resistances.

The gap between experiments and simulations is generally attributed to the presence in fabricated samples of electron-phonon interactions, interface scattering, and surface optical phonons that are typically ignored in QT calculations. Including these effects allows to reproduce the measured electron mobility of monolayer MoS$_2$, but requires well-parameterized scattering rates~[6]. Moreover, recent studies found that the atomic configuration of the amorphous dielectrics surrounding TMDCs plays a crucial role, and highlighted the necessity of constructing oxides at atomistic resolution to obtain accurate mobilities and ``I-V'' characteristics~[7,8]. These \textit{ab initio} investigations combining non-equilibrium Green's functions (NEGF) and density functional theory (DFT) found that geometrical and chemical fluctuations at the oxide interface with TMDCs induce electrostatic potential variations against which electrons scatter. 

In such DFT+NEGF simulations, the computational bottleneck does not reside in the QT part, but in the preparation of their \textit{ab initio} ingredients, i.e., the device's atomic lattice and its Hamiltonian matrix $\mathbf{H}$ that enters the NEGF equations. The $\mathcal{O}(N^3)$ algorithmic scaling of DFT, $N$ being the number of atoms, severely constrains the sizes that can be treated. To circumvent this issue, one possibility is to tile a relatively small unit cell that DFT tools such as CP2K~[9] can handle, and then reconstruct from it the full device structure and its $\mathbf{H}$ matrix. However, this approach, followed by [7,8], yields periodic systems with artificially repeated interface motifs and uncontrollable oxide defect densities. 

Here, we address this system-size limitation by taking advantage of the latest developments in machine learning (ML) representations of atomic systems. Over the last 10 years, molecular dynamics (MD) simulations based on ML interatomic potentials (MLIPs) have been replacing computationally more intensive \textit{ab initio} MD, offering near-DFT accuracy~[10] and enabling the production of device-scale atomic geometries. Similarly, ML models for learning electronic structures have started to appear, initially for molecules and ordered materials with up to a few hundred atoms~[11,12]. After training, they can produce the Hamiltonian matrix corresponding to a given set of atomic coordinates, with DFT-like precision but $\mathcal{O}(N)$ complexity, potentially bypassing the need for intensive \textit{ab initio} electronic-structure calculations. In this context, we introduce a new ML-based Hamiltonian model capable of learning and predicting $\mathbf{H}$ for atomically resolved material stacks, enabling QT simulation of MoS$_2$/HfO$_2$ and MoS$_2$/Al$_2$O$_3$ devices comprising $>$20,000 atoms with dimensions comparable to experiments. This model (i) allows us to reproduce key DFT material (bandstructure) and device (transmission function) properties at a fraction of the computational cost (10,000$\times$ speedups), (ii) scales to larger $N$ without loss of accuracy, and (iii) provides through detailed trajectory analyses insight into the strong interplay between surface oxide atoms, in particular undercoordinated Hf and Al, and the current flowing through MoS$_2$.

\section{\textbf{Methods}}
In DFT+NEGF simulations, the Hamiltonian matrix $\mathbf{H}$ is typically built from atomic orbital interactions expanded in a localized basis of spherical harmonics. To `learn' instead of computing these orbital-interaction terms across varied atomic configurations and device-scale geometries, we have developed \textit{MALOQ}, an SO(2)-equivariant graph neural network operating in the same spherical tensor representations as $\mathbf{H}$. Multiple graph convolution layers (GCL) are used to train shared representations of all nodes and edges of a given atomic graph, which are used to predict the diagonal ([$\mathbf{H_{ii}}$]) and off-diagonal ([$\mathbf{H_{ij}}$]) blocks of $\mathbf{H}$, respectively. Each GCL is a composition of learnable functions that commute with generalized rotations, preserving the rotational symmetries of the atomic lattice. 

\begin{figure*}
\centering\includegraphics[width=\linewidth]{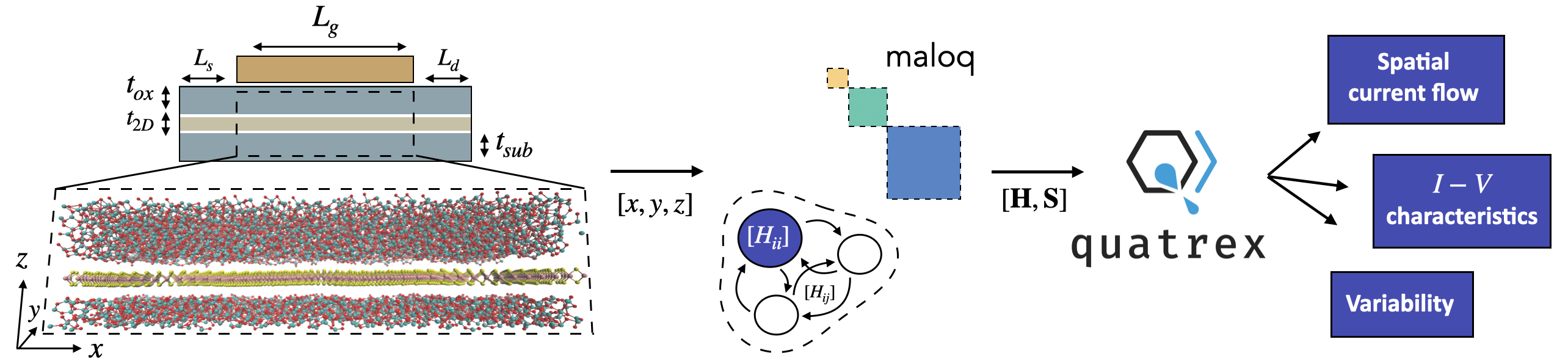}\par
  \caption{\textbf{Overview of the `atomic structure ($\mathbf{xyz}$) $\rightarrow$ learned $\mathbf{H}$ $\rightarrow$ quantum transport' workflow.} The relevant transport regions, a monolayer MoS$_2$ placed between oxide layers, here Al$_2$O$_3$, are first assembled, atom by atom, and placed within a single-gate transistor architecture with a gate length $L_g$, source and drain extensions $L_s$ and $L_d$, oxide thickness $t_{ox}$, and substrate height $t_{sub}$. The amorphous oxide layers are created via a standard melt-and-quench procedure using molecular dynamics with machine-learned interatomic potentials as inputs, before being stacked with MoS$_2$ and undergoing a geometry relaxation. The resulting set of atomic coordinates is passed to the \textit{MALOQ} package, which builds node- and edge-wise embeddings through SO(2)-equivariant graph convolutions. These embeddings are converted into diagonal/off-diagonal blocks [$\mathbf{H_{ii}}$]/[$\mathbf{H_{ij}}$], which are used to reconstruct the Hamiltonian matrix $\mathbf{H_{ML}}$ of the device of interest. Only $\mathbf{H_{ML}}$ is learned, as $\mathbf{S}$ can be determined analytically from the orbital basis. The $\mathbf{H_{ML}}$ and $\mathbf{S}$ matrices serve as inputs to \textit{QuaTrEx}, a quantum transport solver implementing the non-equilibrium Green's function equations. The output quantities are the $I-V$ characteristics and the bond-resolved electronic current.}
\label{fig:1}
\end{figure*}

\textbf{Figure~\ref{fig:1}} presents our complete \textit{ab initio}, ML-driven modeling framework. We first construct our simulation domains atom-by-atom, considering both the channel (here, MoS$_2$) and surrounding oxide layers (here, HfO$_2$ or Al$_2$O$_3$). Their electronic structure is then obtained from \textit{MALOQ}, which predicts a DFT-level Hamiltonian matrix $\mathbf{H}$. This matrix serves as input to a QT solver, \textit{QuaTrEx}~[13], that returns non-equilibrium charge densities, electrostatic potentials, ``I-V'' characteristics, and spatial current distributions based on the NEGF formalism.

\section{\textbf{Results}}

\subsubsection{Dataset generation \& training} 
\begin{figure*}
\centering\includegraphics[width=\linewidth]{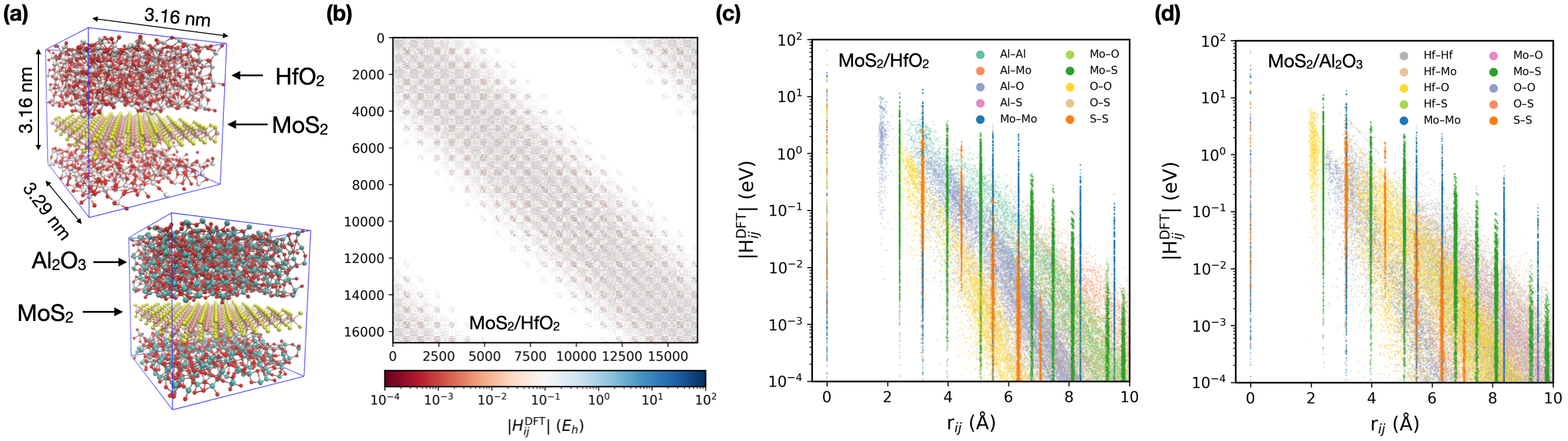}\par
  \caption{\textbf{DFT-level dataset generation for MoS$_2$/oxide electronic structures.} \textbf{(a)} Atomic samples selected from the generated training dataset, which consists of a total of 50 $\sim$2,000-atom 2D/oxide structures. The amorphous oxides were all created with an MD melt-quench process: melting for 20 ps at 4000 K, and quenching for 400 ps in a constant ramp down to 300 K. The oxide is then cleaved, sandwiched with the 2D monolayer, annealed for 20 ps at 1000 K with DFT D3 corrections, and finally relaxed. \textbf{(b)} Heatmap of the Hamiltonian matrix $\mathbf{H}$ for one of these MoS$_2$/HfO$_2$ structures, as computed with CP2K using a DZVP basis for MoS$_2$ and a SZV basis for the oxides, resulting in a total size of 16,608$\times$16,608 and 138 million nonzero electronic orbital interactions. Each of these matrices undergoes a decomposition into [$\mathbf{H_{ii}}$]/[$\mathbf{H_{ij}}$] sub-matrices, followed by an analytical basis transformation into a space of uncoupled spherical tensors. \textit{MALOQ} is trained to minimize the $L2+\sqrt{L2}$ loss over these tensor elements up to an interatomic distance of 10 \AA. \textbf{(c)} 1 million randomly sub-sampled nonzero matrix elements of $\mathbf{H}$ ($|H_{ij}|$) from the 138 million in (b), as a function of the interatomic distance ($r_{ij}$) for one MoS$_2$/HfO$_2$ sample. Each scatter point is colored according to the atomic element interaction it originates from in the Hamiltonian. The regular vertical patterns in the scatter points indicate Mo-S interactions, while the scattered background contains the orbital interactions within the oxide and at the interfaces with MoS$_2$. We note that a minimum cell dimension of 2 nm (corresponding to a distance of 10 \AA\;) between periodic images is necessary to fully saturate the decay of $\mathbf{H_{ij}}$ interaction terms and avoid artefacts stemming from the periodic boundary conditions of DFT. \textbf{(d)} Same as (c), but for MoS$_2$/Al$_2$O$_3$.}
\label{fig:2}
\end{figure*}

To produce Hamiltonian matrices for large-scale MoS$_2$/HfO$_2$ and MoS$_2$/Al$_2$O$_3$ systems, \textit{MALOQ} must first be trained on the $\mathbf{H_{DFT}}$ of structures including Mo, S, Al, Hf, and O atoms. We create them by extracting periodic snapshots along high-temperature MD trajectories of 3.2$\times$3.2$\times$2.8 nm$^3$ MoS$_2$/oxide samples performed with LAMMPS~[14] parameterized with MACE-MH1~[10]. Two such snapshots are pictured in \textbf{Fig.~\ref{fig:2}(a)}. Their Hamiltonian $\mathbf{H_{DFT}}$ and Overlap $\mathbf{S_{DFT}}$ matrices are computed with CP2K~[9], using a double-$\zeta$ valence polarized (DZVP) basis for MoS$_2$, and a single-$\zeta$ (SZV) basis for the oxides. The full training dataset contains 50 structures, 115,200 atoms, and $\sim$6.9B orbital-interaction coefficients that are retained up to an interatomic cutoff distance of 10~\AA\; (\textbf{Fig.~\ref{fig:2}(c)-(d)}).

\subsubsection{Model Validation}  
\begin{figure*}
  \centering\includegraphics[width=\linewidth]{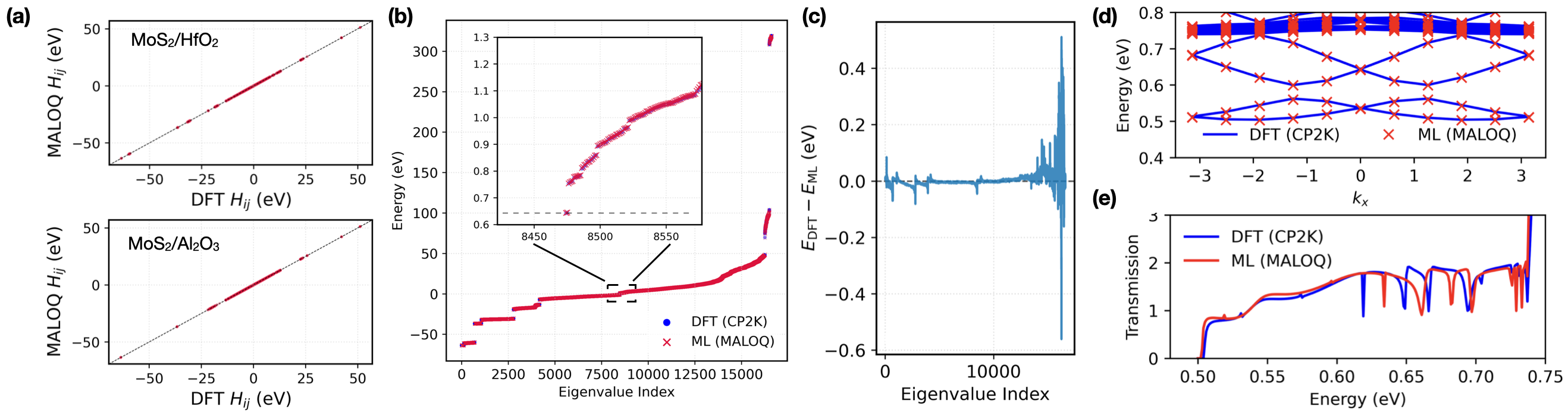}\par
  \caption{\textbf{\textit{MALOQ} training validation.} \textbf{(a)} Parity plots between the DFT and machine-learned matrix elements of $\mathbf{H}$, showing perfect alignment for both MoS$_2$/HfO$_2$ (top) and MoS$_2$/Al$_2$O$_3$ structures, with an average matrix element error of 39 $\mu$eV. \textbf{(b)} Comparison between all eigenvalues (16,608) of $\mathbf{H_{DFT}}$ (blue circles) and $\mathbf{H_{ML}}$ (red crosses) for the MoS$_2$/HfO$_2$ system made of 2,088 atoms. The inset zooms into the edge of the MoS$_2$ conduction band. \textbf{(c)} Energy difference (error) between all eigenvalues of $\mathbf{H_{DFT}}$ and $\mathbf{H_{ML}}$ in \textbf{(b)}. On average, the error is 38 meV, but it is largely concentrated at high energies, outside the range of interest. \textbf{(d)} Conduction bandstructure of the MoS$_2$/HfO$_2$ sample from (a)-(c), as obtained from CP2K's (solid blue lines) and \textit{MALOQ}'s (red crosses) Hamiltonian matrix. \textbf{(e)} Transmission function through the same MoS$_2$/HfO$_2$ structure as in \textbf{(d)} computed with \textit{QuaTrEx} and using $\mathbf{H_{DFT}}$ (blue line) and $\mathbf{H_{ML}}$ (red line) as inputs. Note that pure MoS$_2$ extensions are added on each side of the MoS$_2$/HfO$_2$ central part and act as electron injection contacts.}
\label{fig:3}
\end{figure*}

In \textbf{Fig.~\ref{fig:3}}, we demonstrate \textit{MALOQ}'s accuracy in learning $\mathbf{H}$ for the systems of interest. The predicted ($\mathbf{H_{ML}}$) entries of two 3.2$\times$3.2$\times$2.1 nm$^3$ MoS$_2$/oxide systems consisting of 2,088 and 2,520 atoms show excellent agreement with their DFT labels (\textbf{Fig.~\ref{fig:3}(a)}). Diagonalizing $\mathbf{H_{ML}}$ for MoS$_2$/HfO$_2$ results in nearly exact eigenvalues around the MoS$_2$ conduction band edge (\textbf{Fig.~\ref{fig:3}(b)}), with an average error of 38 meV across all 16,608 eigenvalues. This remaining error is mostly distributed in the highest unoccupied states (\textbf{Fig.~\ref{fig:3}(c)}) that do not contribute to transport. With this accuracy of $\mathbf{H_{ML}}$, we can very well reproduce the DFT-computed conduction bandstructure (\textbf{Fig.~\ref{fig:3}(d)}) and transmission function (\textbf{Fig.~\ref{fig:3}(e)}). 

\subsubsection{Generalization at scale} 
\begin{figure*}
  \centering\includegraphics[width=0.95\linewidth]{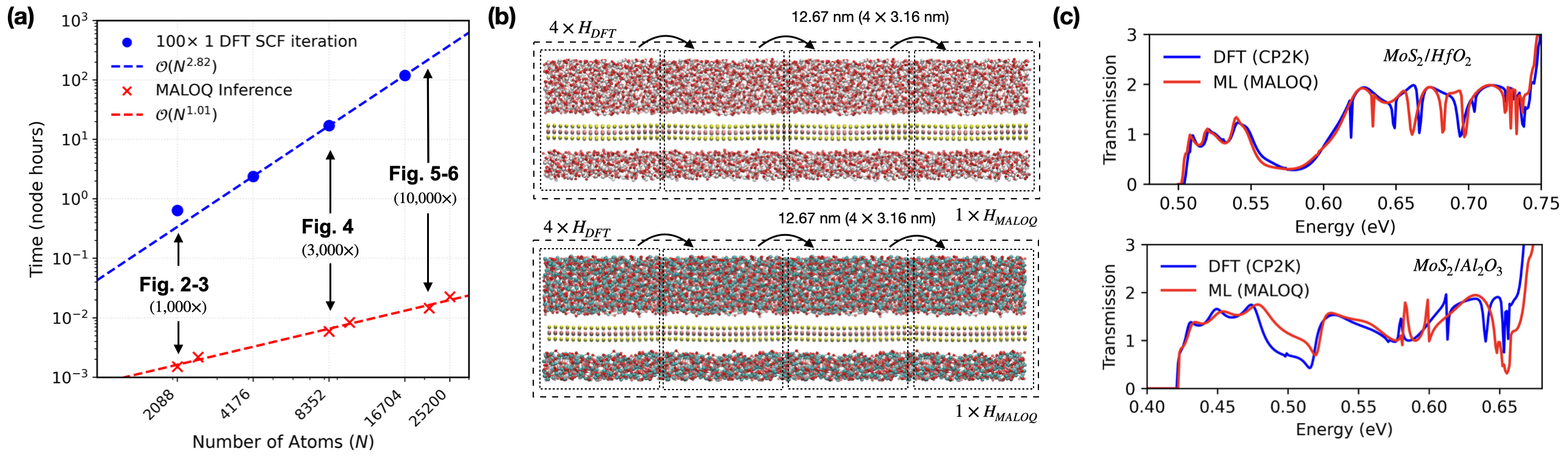}\par
  \caption{\textbf{Computational complexity and generalization of \textit{MALOQ}.} \textbf{(a)} Computational cost in node hours of CP2K (blue circles) and \textit{MALOQ} (red crosses) to produce the Hamiltonian matrix of systems made of a different number of atoms, from 2,088 to 16,704, on GH200 GPUs. The DFT calculations converge on average within 100 self-consistent field (SCF) iterations, while the ML approach is single-shot. The dashed lines serve as visual guides and show a measured time complexity of $\mathcal{O}(N^{2.82})$ for CP2K and $\mathcal{O}(N^{1.01})$ for \textit{MALOQ}. The speedups brought by \textit{MALOQ} with respect to DFT are indicated as double-arrows, with a reference to the figures using the annotated sizes. \textbf{(b)} 12.8$\times$3.2$\times$2.8 nm$^3$ MoS$_2$/HfO$_2$ and MoS$_2$/Al$_2$O$_3$ structures consisting of the repetition of four identical 3.2$\times$3.2$\times$2.8 nm$^3$ cells, containing $\sim$8,000 atoms. \textbf{(c)} Transmission function through the structures in (b), near their conduction band edges. \textit{QuaTrEx} was used to compute all curves, with inputs from CP2K (blue line) and \textit{MALOQ} (red line). Importantly, CP2K only returned the Hamiltonian matrix of the small cell, which was then tiled, while \textit{MALOQ} received the whole system at once, without knowledge of its internal periodicity.} 
  \label{fig:4}
\end{figure*}

While DFT is fast enough to obtain the $\mathbf{H}$ matrices of the training structures in \textbf{Fig.~\ref{fig:2}(a)}, its poor computational scalability prevents the modeling of device-scale systems within reasonable times. To quantify this and underline the benefit of ML approaches, we report in \textbf{Fig.~\ref{fig:4}(a)} the computational cost stemming from the generation of $\mathbf{H}$ with either CP2K or \textit{MALOQ} for 2,088 to 16,704 atoms. Assuming convergence within 100 self-consistent field (SCF) iterations in DFT, the simulation of the largest structure takes $\sim$200 node hours (NHs) (each node has 4 GH200 GPUs), and the measured time complexity scales with $\mathcal{O}(N^{2.82})$, close to its algorithmic complexity of $\mathcal{O}(N^{3})$. In contrast, for the same size, \textit{MALOQ}'s linear-scaling inference produces $\mathbf{H}$ in $<$0.1 NH, a $>$10,000$\times$ speedup over DFT. A single inference on a 25k-atom system (0.1 instead of 200 NHs) thus fully amortizes the computational cost of our dataset generation (30 NHs) and training (120 NHs). 

\textbf{Figure~\ref{fig:4}(b-c)} illustrates the ability of \textit{MALOQ} to generalize $\mathbf{H}$ to larger, unseen MoS$_2$/oxide structures of size 12.8$\times$3.2$\times$2.8 nm$^3$ while preserving its prediction accuracy with respect to DFT. To minimize the computational burden of CP2K, the selected systems consist of the fourfold repetition along the $x$-axis of the same 3.2$\times$3.2$\times$2.8 nm$^3$ unit cell. The reference $\mathbf{H_{DFT}}$ can thus be produced by directly tiling a smaller matrix, while \textit{MALOQ} receives the input structure as a whole. For both HfO$_2$ and Al$_2$O$_3$, the transmission functions, as computed with \textit{QuaTrEx} using $\mathbf{H_{DFT}}$ and $\mathbf{H_{ML}}$ as inputs, are in very good agreement over a large energy range.

\subsubsection{Device simulations} 
\begin{figure*}\centering\includegraphics[width=0.90\linewidth]{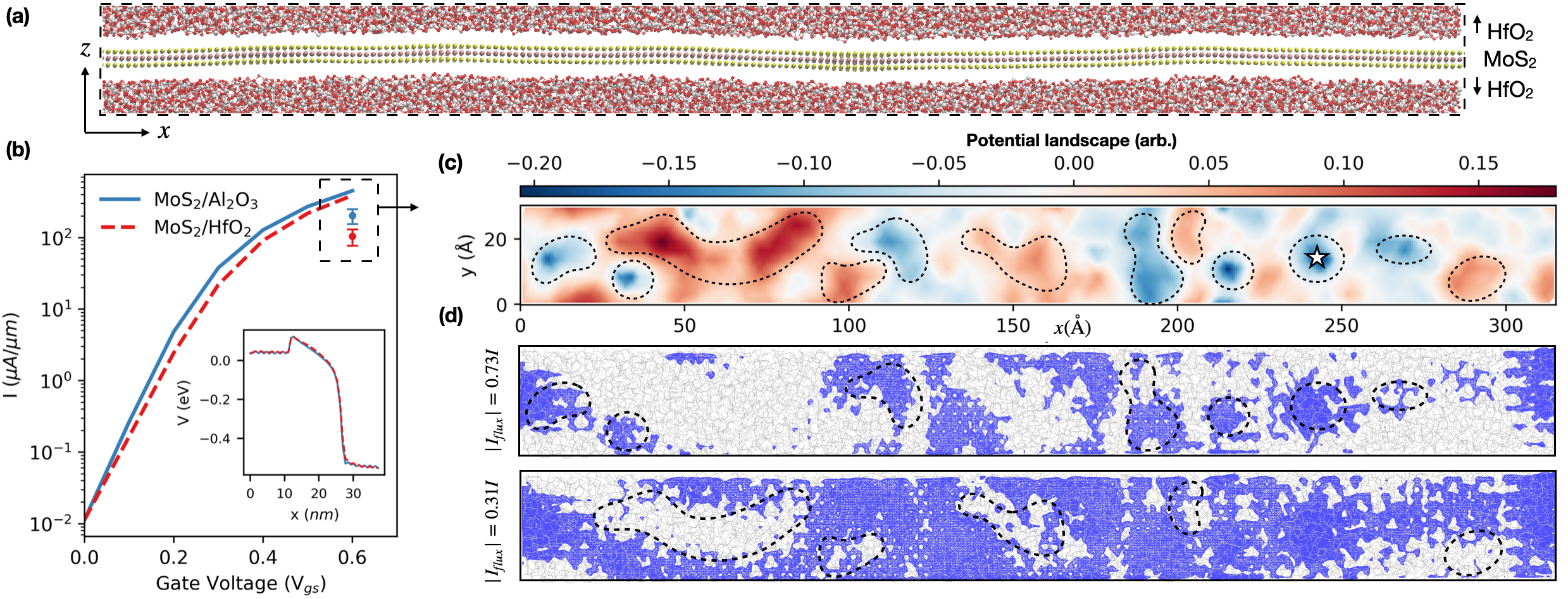}\par
  \caption{\textbf{Current flow analysis in MoS$_2$/oxide devices.} \textbf{(a)} Top view ($x$-$z$ plane) of an MoS$_2$/HfO$_2$ atomic structure inserted into the $n$-type FET architecture of Fig.~\ref{fig:1} with $L_g$=15 nm, $L_s$=$L_d$=11.5 nm (total length $L_{ch}$=38 nm), and $t_{sub}$=20 nm. The source and drain are doped with a donor concentration $N_D$=5e13 cm$^{-2}$. The equivalent oxide thickness (EOT) is set to 0.8 nm ($t_{ox}$=4 nm for HfO$_2$ with $\epsilon_R$=20 or $t_{ox}$=2 nm for Al$_2$O$_3$ with $\epsilon_R$=10). \textbf{(b)} Transfer characteristics of MoS$_2$/HfO$_2$ and MoS$_2$/Al$_2$O$_3$ FETs with the specifications in (a) at $V_{ds}$=0.6 V. The solid lines refer to QT simulations where only the MoS$_2$ channel is part of the transport domain and the dielectric layers are treated as ideal insulators. The total number of atoms is 2,160, with the size of $\mathbf{H_{DFT}}$ equal to 37,440. The symbols (error bars) at $V_{gs}$=0.6 V are the averages (standard deviations) of three QT calculations per device type, explicitly including the oxides in the NEGF equations ($>$20,000 atoms, size of $\mathbf{H_{ML}}$ $>$180,000). They were computed at $V_{gs}$=0.6 V using the electrostatic potential profiles shown as inset. \textbf{(c)} Top view ($x$-$y$ plane) of the electrostatic potential landscape experienced by the MoS$_2$ atomic plane for the structure in (a) under flat band conditions. The formation of hills (red) and valleys (blue) can be observed. \textbf{(d)} Top view of current isosurfaces at a high (top) and low (bottom) value. In the upper plot, current hot spots are clearly visible that correlate with the presence of potential valleys in (c). In the lower one, regions with no current align well with hill's locations.} 
  \label{fig:5}
\end{figure*}

After establishing the prediction and scaling capabilities of \textit{MALOQ}, we leverage them to produce the $\mathbf{H_{ML}}$ of MoS$_2$/Al$_2$O$_3$ and MoS$_2$/HfO$_2$ single-gate $n$-type FETs with the architecture of \textbf{Fig.~\ref{fig:1}} and the channel configuration of \textbf{Fig.~\ref{fig:5}(a)}. Each structure is more than 30 nm long, contains $>$20,000 atoms, with the size of $\mathbf{H_{ML}}$ exceeding 180,000. We first compare in \textbf{Fig.~\ref{fig:5}(b)} the ``ON-state'' currents of these FETs at $V_{ds}$=$V_{gs}$=0.6 V and fixed EOT of 0.8 nm, as computed with \textit{QuaTrEx} with and without explicitly accounting for the atomic granularity and amorphous nature of the surrounding oxides. \textit{The presence of a physical oxide reduces the current by a factor of 2 (Al$_2$O$_3$) to 3 (HfO$_2$).} 

Refs.~[7,8] attributed this reduction to the oxide-induced formation of electrostatic hills and valleys, also present in \textbf{Fig.~\ref{fig:5}(c)}, that affect the current trajectories. As our simulation domains are free from the artefacts of periodicity in these studies and exhibit a larger variety of structural motifs, we can better visualize how potential variations impact the current flow through, for example, a MoS$_2$/HfO$_2$ device. In particular, current isosurface plots (\textbf{Fig.~\ref{fig:5}(d)}) show that the current crowds at the location of potential valleys. The electrostatic influence of the oxide thus narrows the effective channel width compared to the case with an ideal dielectric.

\begin{figure*}\centering\includegraphics[width=0.95\linewidth]{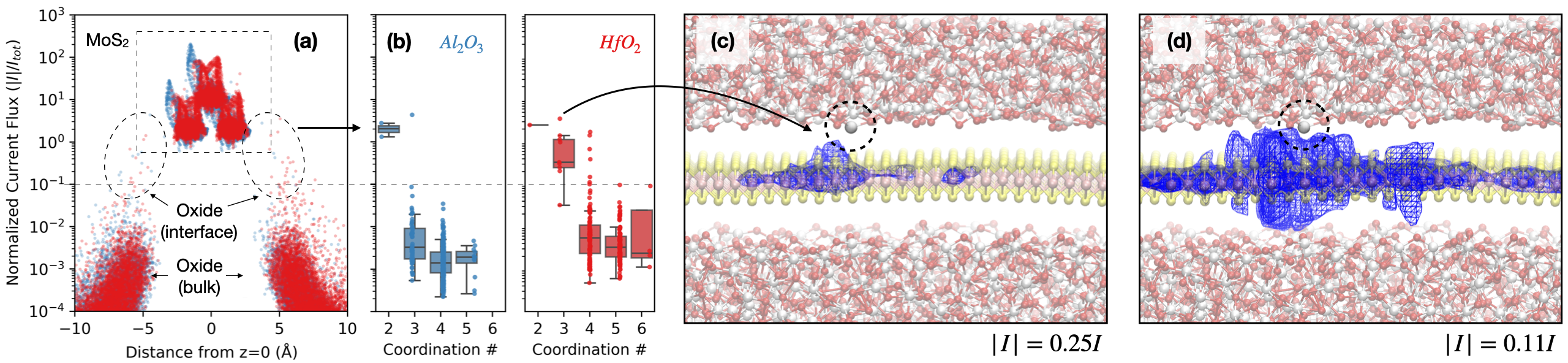}\par
  \caption{\textbf{Scattering at oxide interfaces} \textbf{(a)} Current flux passing through each atom of the three MoS$_2$/HfO$_2$ (red dots) and three MoS$_2$/Al$_2$O$_3$ (blue dots) structures used in Fig.~\ref{fig:5}(b), with $z$=0 corresponding to the average location of the Mo plane along $x$. The dashed box indicates the flux through the Mo and S atoms. Lateral variation in the current flux through MoS$_2$ along $z$ is from macroscopic bending of each monolayer. The dashed ovals highlight large current fluxes through interface oxide atoms. \textbf{(b)} Same as in (a), but for the flux through the Hf and Al atoms in close contact with MoS$_2$, sorted as a function of the metal atom's oxygen coordination number. \textbf{(c)} Isosurface of the current flowing through the MoS$_2$/HfO$_2$ structure in Fig.~\ref{fig:5}(c), at the region marked with a star, with focus on the trajectory deformation induced by a surface undercoordinated Hf atom. \textbf{(d)} Same as in (c), but for an isosurface taken at a lower current value.} 
  \label{fig:6}
\end{figure*}

To probe the atomistic origins of these potential valleys, we first report in \textbf{Fig.~\ref{fig:6}(a)} the current flux through each atom of six independent MoS$_2$/oxide samples, after normalizing their total current. As expected, most of the current flows through the MoS$_2$ channels, especially through their Mo planes, and the current magnitude exponentially decreases away from them. Nevertheless, significant leakage currents can be observed at the semiconductor-oxide interfaces. Each of these leakage points can be traced back to the presence of either Hf or Al undercoordinated metal atoms (\textbf{Fig.~\ref{fig:6}(b)}), which affect the electron propagation. For example, the location marked with a star in \textbf{Fig.~\ref{fig:5}(c)} corresponds to a potential valley. Close examination reveals its close proximity with an undercoordinated Hf atom, current crowding within the MoS$_2$ channel below (\textbf{Fig.~\ref{fig:6}(c)}) as well as localized leakage into the oxide (\textbf{Fig.~\ref{fig:6}(d)}).

\section{\textbf{Conclusion and Outlook}}
Predicting the electronic structure of large-scale atomic systems with \textit{MALOQ} unlocks unprecedented levels of insight into how electronic current flows through MoS$_2$/oxide interfaces with atomistic granularity, and how this affects the performance of the resulting nanoscale FETs. Our workflow is ready to be deployed across arbitrary atomic elements and combined with other scattering mechanisms for the computational design of next-generation nanoelectronic devices and interconnects.

\vspace{0.2cm}

\fontsize{9}{9}\selectfont{
\noindent\textbf{Acknowledgments:} The authors acknowledge MARVEL, Swiss Chips, and the Swiss National Supercomputing Center (CSCS). 

\vspace{0.2cm}

\noindent\textbf{References:}
[1] Z.~Zhu et al., IEDM, 10.4.1-10.4.4 (2025).
[2] J.~Kwon et al., Nat. Mat. 25, 824 (2025).
[3] T.~Pe\~na et al., Nat. Nano. 21, 803 (2026).
[4] P.-C. Shen et al., Nature 593, 211 (2021).
[5] Y.~Zhao et al., J. of Phys. Chem. C 126, 12100 (2022).
[6] M.~Hosseini et al., IEEE Trans. on Elec. Dev. 62, 3192 (2015).
[7] M.~Dossena et al., npj 2D Mat. and Appl. 9, 67 (2025).
[8] F.~Ducry et al., npj 2D Mat. and Appl. 10, 49 (2026).
[9] T.~D.~K\"uhne et al., J. of Chem. Phys. 152, 194103 (2020).
[10] I.~Batatia et al., NeurIPS 35, 11423 (2022).
[11] X.~Gong et al., Nat. Comm. 14, 2848 (2023).
[12] M.~Kaniselvan et al., 14th ICLR (2026).
[13] N.~Vetsch et al., SC25 (2025).
[14] A.~P.~Thompson et al., Comp. Phys. Comm. 271, 108171 (2026).
}


\newpage

\end{document}